\documentclass[12pt]{elsarticle}
\usepackage{amssymb}
\usepackage{amssymb,amsmath}
\biboptions{numbers,sort&compress}
\usepackage{mathtools}
\usepackage{placeins}
\usepackage{eqnarray}
\usepackage{multirow}
\usepackage{booktabs} \renewcommand{\arraystretch}{1}
\usepackage{float}
\usepackage[left=2cm, right=2cm, top=2cm, bottom  =2cm]{geometry}

\usepackage{graphicx}
\usepackage{subfig}

\usepackage{bm}
\usepackage[colorlinks,bookmarksopen,bookmarksnumbered,citecolor=red,urlcolor=red]{hyperref}\usepackage{soul}
\usepackage{enumerate}
\usepackage{enumitem}
\usepackage[nodisplayskipstretch]{setspace}
\usepackage{textcomp}
\usepackage{graphicx} 
\usepackage{todonotes}

\graphicspath{{Figures/}}
\usepackage{txfonts}
\usepackage{dsfont}
\newlist{steps}{enumerate}{1}
\setlist[steps, 1]{label = \textbf{Step \arabic*:},leftmargin=1.3cm}

\newcommand{\Hf}{ \mathcal{H}} 
\newcommand{\Hfit}{ \widetilde{\mathcal{H}}(s)}
\newcommand{\Hr}{ \widetilde{\mathcal{H}}}
\newcommand{\norm}[1]{\left\lVert#1\right\rVert}
\newcommand{\ssAr}{ \widetilde{\varmathbb{A}}}
\newcommand{\ssBr}{ \widetilde{\varmathbb{B}}}
\newcommand{\ssCr}{ \widetilde{\varmathbb{C}}}

\newcommand{\relerr}{\mathcal{E}_{L_2}^{rel}}

\newcommand{\frf}{{\small{\textsf{FRF}}}}
\newcommand{\frfs}{{\small{\textsf{FRF}s}}}
\newcommand{\vf}{{\small{\textsf{VF}}}}
\newcommand{\ls}{{\small{\textsf{LS}}}}

\journal{~}
    \usepackage[symbol]{footmisc}

\begin{document}

\begin{frontmatter}

\title{Estimating Dispersion Curves from Frequency Response Functions via Vector-Fitting}

\author[Mech1]{Mohammad I. Albakri\
\fnref{firstfoot}} \corref{note1}
\author[Mech2]{Vijaya V. N. Sriram Malladi\fnref{firstfoot}}
\author[Math]{Serkan Gugercin}
\author[Mech]{Pablo A. Tarazaga}

\address[Mech1]{Department of Mechanical Engineering, Tennessee Technological University}
\address[Mech2]{Department of Mechanical Engineering - Engineering Mechanics, Michigan Technological University}

\address[Mech]{Department of Mechanical Engineering, Virginia Polytechnic Institute and State University}
\address[Math]{Department of Mathematics, Virginia Polytechnic Institute and State University}
\cortext[note1]{Address all correspondence to this author. Email: malbakri@vt.edu}
\fntext[firstfoot]{These authors have contributed equally}

\begin{abstract}
{\it 
Driven by the need for describing and understanding wave propagation in structural materials and components, several analytical, numerical, and experimental techniques have been developed to obtain dispersion curves.  Accurate characterization of the structure (waveguide) under test is needed for analytical and numerical approaches. Experimental approaches, on the other hand, rely on analyzing waveforms as they propagate along the structure. Material inhomogeneity, reflections from boundaries, and the physical dimensions of the structure under test limit the frequency range over which dispersion curves can be measured.  

In this work, a new data-driven modeling approach for estimating dispersion curves is developed. This approach utilizes the relatively easy-to-measure,  steady-state Frequency Response Functions (\frfs) to develop a state-space dynamical model of the structure under test. The developed model is then used to study the transient response of the structure and estimate its dispersion curves. This paper lays down the foundation of this approach and demonstrates its capabilities on a one-dimensional homogeneous beam using numerically calculated \frfs. Both in-plane and out-of-plane \frfs~corresponding, respectively, to longitudinal (the first symmetric) and  flexural (the first anti-symmetric) wave modes are analyzed. The effects of boundary conditions on the performance of this approach are also addressed.\\
}
\end{abstract}


\end{frontmatter}

\section{Introduction}
Understanding wave propagation in structures is essential for numerous applications such as structural health monitoring, material characterization, stress-state identification \cite{Stress2018}, event detection and localization \cite{schloemann2015vibration,poston2015towards, woolard2017applications}, vibration suppression, and elastic meta-structures \cite{caloz2011metamaterial}. Key to this understanding is the ability to describe the frequency-dependent nature of wave propagation characteristics governed by dispersion curves. Dispersion relations of elastic waves are determined by geometric and material characteristics of the waveguide, as well as the nature of the deformations induced by the propagating wave (the wave mode) \cite{graff2012wave}.

Analytical models have been developed to derive dispersion relations for numerous materials and waveguides \cite{graff2012wave}. For commonly used homogeneous and heterogeneous materials, dispersion relations have been documented in various handbooks and textbooks \cite{chaigne2014structural, liu2001elastic}. {Numerical methods, such as time and frequency domain spectral element method \cite{Doyle, SEM:gopalakrishnan2007, SEM:lee2009, 2018SEM_Rev}, the semi-analytical finite element method \cite{SAFE:2008, SAFE_GMM2019}, the wave finite element method \cite{mace2008WFE, zhou2015wave}, and transfer function and transfer matrix methods \cite{2017_TM1, 2017_TM2} have also been developed to describe wave propagation along uniform and periodically varying waveguides}. For realistic wave propagation characteristics to be obtained, accurate description of material and geometric characteristics of the waveguide is required \cite{bilbao2004wave}. Accurate characterization of waveguides can be challenging, especially when inhomogeneous materials, such as wood and steel-reinforced concrete, and unconventional structures, such as 3D printed and smart structures with integrated sensors and actuators, are being considered. Experimental methods for measuring dispersion relations have also been developed where a tone burst is induced in the structure under test and then analyzed as it propagates along the structure. However, material inhomogeneity, complex boundary conditions, and the physical dimensions of the structure under test pose challenges to such techniques. In particular, the ability of current techniques to measure the low frequency portions of dispersion curves, which are very important for event localization and stress-state identification applications, is limited by structure's dimensions, damping, and reflections from structure's boundaries.

Data-driven modeling provides an alternative approach to create numerical models that capture the dynamic response of the structure under test. Such models rely completely on measured responses, thus, any knowledge of the underlying physical characteristics of the structure is not required.
In this context, the data amounts to a large set of frequency response measurements and the modeling amounts to constructing a low-dimensional rational function to fit this data in an appropriate sense. This is a heavily studied topic with varying approaches based on the eventual goal;
{see, e.g.,
\cite{aca90,ALI18,mayo2007fsg,schulze2018data} for rational approximations  that enforce interpolation of the measured data; 
\cite{hokanson2017projected,gonnet2011robust,gustavsen1999rational,Drmac-Gugercin-Beattie:VF-2014-SISC,berljafa2017rkfit}
for rational approximations that minimize a least-squares
deviation from the measured data; and 
\cite{nakatsukasa2018aaa,nakatsukasa2019algorithm} for rational approximations  that combine the interpolation and least-squares formulation.} Here, the least-squares based Vector Fitting \cite{gustavsen1999rational} framework will be employed.

In this work, a new data-driven approach for estimating dispersion curves is developed. Steady-state dynamic responses, in the form of Frequency Response Functions (\frfs), are used to build the data-driven model of the structure under consideration. The developed model is then used to simulate the transient dynamic response of the structure which, in turn, is analyzed to reconstruct dispersion curves. The capabilities of this approach are demonstrated on a simple one-dimensional structure for which exact dispersion curves are attainable. The work builds on a previous effort by the authors \cite{Data2018} and discusses in detail the development of this approach using numerical experiments. The applicability of this approach to symmetric (longitudinal) and anti-symmetric (flexural) wave modes is investigated. {These wave modes are selected as they are the simplest to analyze theoretically and experimentally. Other wave modes along with wave propagation in two-dimensional structures, such as plates, will be addressed in future studies.} This new approach is proposed as an alternative to the traditional experimental techniques, with the potential of overcoming the aforementioned limitations, especially when low-frequency portions of the dispersion curves are of interest. Furthermore, the fact that the developed approach relies on the steady-state dynamic response to estimate dispersion curves, as opposed to transient waveform measurements, relaxes sampling rate requirements and improves signal-to-noise ratio in testing.

This paper is structured as follows: Section 2 briefly presents the development of the numerical model used to generate the \frfs~to construct data-driven models. Elementary rod and Timoshenko beam approximate theories, along with the frequency domain spectral element method, are used to develop this model, which also provides the exact dispersion curves for comparison.  Data-driven rational approximation is discussed in Section 3 where the simulated \frfs~are used to build a data-driven, state-space model that describes the dynamics of the beam under test. The layout of the newly developed approach is discussed in Section 4 where dispersion curves of the first anti-symmetric wave mode are estimated. The applicability of this approach to estimate dispersion relations for symmetric wave modes are presented in Section 5. Section 6 discusses the effects of boundary conditions on the performance of this approach. Finally, concluding remarks are presented in Section 7.  

\section{Mathematical Model and Spectral Element Formulation}

\begin{figure}[ht]
\centering
\includegraphics[trim={4cm 7cm 4cm 3.5cm},clip, width=0.75\textwidth]{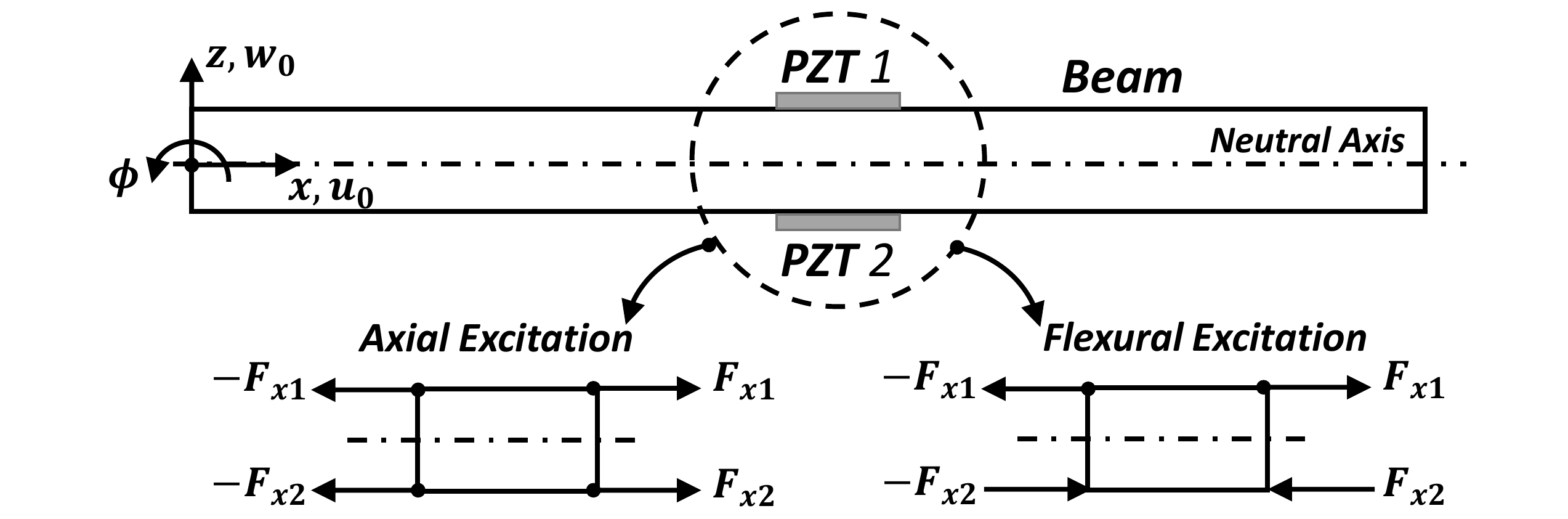}
\caption{Schematic of a beam excited with surface mounted piezoelectric actuators (PZT).}
\label{fig:beam}
\end{figure}

Consider a long beam excited with a pair of identical piezoelectric actuators, as shown in Figure \ref{fig:beam}. The piezoelectric actuators can be excited either in-phase to generate pure longitudinal deformations or out-of-phase for pure flexural excitation. This excitation configuration is selected here since it is commonly used in experimental studies. Due to the structure symmetry around its neutral axis, longitudinal and flexural deformations are completely uncoupled. For simplicity, the piezoelectric actuators are replaced with two pairs of longitudinal forces acting at the beam's upper and lower surfaces, as shown in Figure \ref{fig:beam}. {Readers are referred to \cite{lee2013dynamic,albakri2017dynamic} for detailed formulation of piezoelectric-augmented structures.}
\vspace{-0cm}

The mass and stiffness of the piezoelectric actuators are assumed to be negligible compared to those of the beam. Following the Elementary rod and the Timoshenko beam approximate theories, the elastodynamic equations of motion can be expressed as 
\begin{eqnarray} 
\label{eq:pde}
\rho A \frac{\partial^2u_0}{\partial t^2}-\frac{\partial u_0}{\partial x} \left( EA\frac{\partial u_0}{\partial x}\right ) &=&qA,\nonumber \\
\rho A \frac{\partial^2w_0}{\partial t^2}-GA\bar{K} \frac{\partial}{\partial x}\left ( \frac{\partial w_0}{\partial x}-\phi \right ) &=&P,\nonumber \\ 
\rho I \frac{\partial^2\phi}{\partial t^2}-EI\frac{\partial^2\phi}{\partial x^2}-GA \bar{K} \left ( \frac{\partial w_0}{\partial x}-\phi\right ) &=& 0,
\end{eqnarray}
with the boundary conditions 
\begin{equation}
EA\frac{\partial u_0}{\partial x}=F_x,\hspace{30pt} GA\bar{K} \left ( \frac{\partial w_0}{\partial x}-\phi \right ) = F_z,\hspace{15pt} \mbox{and} \hspace{15pt}  EI\frac{\partial \phi}{\partial x}=M,
\end{equation}
where $u_0$ and $w_0$ denote the longitudinal and lateral displacements of the beam's neutral axis, respectively;  $\phi$ denotes the angle of rotation of the neutral axis normal vector;  $\rho$, $E$, and $G$ are, respectively, the material's volumetric mass density, elasticity modulus, and rigidity modulus; $A$ is the beam's cross sectional area;  and $I$ is its second moment of area. The Timoshenko correction factor, $\bar{K}$, is determined by matching the high-frequency wave speed to that of Rayleigh waves \cite{Doyle}. { $q\left ( x,t \right )$ is the externally applied axial body force per unit volume and $P\left ( x,t \right)$ is the externally applied transverse distributed load. Both $q$ and $P$ are non-existent in the current analysis.} The externally applied concentrated longitudinal forces, lateral forces, and bending moments are denoted by $F_x$, $F_z$, and $M$, respectively. 

The frequency domain Spectral Element Method (SEM) is used to solve the elastodynamic equations of motion and to obtain \frfs~corresponding to longitudinal and flexural deformations. This method is selected over other numerical methods, such as the FEM, due to its superior accuracy \cite{SEM:lee2009}. This is especially important when the high-frequency dynamic response is of interest. The spectral element formulation is briefly presented in this section.  

With the SEM, all variables appearing in the equations of motion, along with boundary conditions, are first transformed to frequency domain using the discrete Fourier transform, and a spectral solution of the following form is assumed
\begin{eqnarray} \label{eq:specsol}
\mathbf{u}(x,t)=\frac{1}{\hat{N}}\sum_{n=0}^{\hat{N}-1} \sum_{m=1}^{M} r_{lm}A_m e^{-\dot{\imath} \left ( k_{mn}x-\omega_nt \right )},
\end{eqnarray}
where $\mathbf{u}\left (x,t \right )=[\begin{array}{ccc} u_0 \left ( x,t \right ) & w_0 \left ( x,t \right ) & \phi \left ( x,t \right ) \end{array}]^T$ is the response variables vector in time domain; $\hat{N}$ is the number of spectral components considered in the discrete Fourier transform, $M$ is the number of wave-modes contributing to the displacement-field,  $k_{mn}$ is the wave number corresponding to the $m^{th}$ mode at the $n^{th}$ angular frequency ($\omega_n$), and $\dot{\imath}=\sqrt{-1}$. Fourier coefficients at $\omega_n$ are represented in terms of the scaling matrix, $\bf r$, and the amplitude vector $\bf A$. The variable $l$ takes the value of $1$, $2$, or $3$ for $u_0 $, $w_0$, and $\phi$, respectively.

Fourier coefficients corresponding to the coupled variables, $w_0$ and $\phi$, can be scaled relative to either one of them. In this work, the coefficient corresponding to $w_0$ is chosen to be the scaling coefficient, hence, $r_{2m}=1,~ \forall m$. The uncoupled axial deformations $u_0$, on the other hand, can be solved for independently. Thus, no scaling factor is required for this component.   

Substituting the spectral solution, \eqref{eq:specsol} into the governing equations \eqref{eq:pde} yields

\begin{equation*}
\left(-\rho \omega_n^2+Ek_{mn}^2 \right)A_m =0,
\end{equation*}
\begin{eqnarray}
 \left[ {\begin{array}{cc}
   GA\bar{K} k_{mn}^2-\rho A \omega_n ^2 & -i GA \bar{K} k_{mn} \\
   i GA \bar{K} k_{mn} & EI k_{mn}^2 - \rho I \omega_n ^2 + GA \bar{K} \\
  \end{array} } \right] \left[ {\begin{array}{cc}
   1\\ r_{3m} \\
  \end{array}} \right]A_m = 0,
\end{eqnarray}
For each frequency of interest, $\omega_n$, the characteristic equations are solved for the wavenumber $k_{mn}$, which results in three wavemodes. These wavemodes represent the propagating first symmetric and first anti-symmetric modes, along with the evanescent second anti-symmetric mode. The later wavemode transfers to propagating upon passing through the cut-off frequency $\omega _c=\sqrt{GA \bar{K} / \rho I}$. Group velocity for each propagating mode is then calculated as $V_{Gm}=\partial \omega_n / \partial k_{mn}$. These results are used in later sections to assess the performance of the data-driven approach. Once the characteristic equations are solved, the scaling constants, $r_{3m}$, are calculated for each mode $m$.

To calculate \frfs~corresponding to longitudinal and flexural deformations of the structure, the dynamic stiffness matrix, $\bf{K}(\dot{\imath}\omega)$ is first evaluated. This matrix relates the nodal displacements vector, $\bf{d}$, and the nodal forces and moments vector, $\bf{F}$, in the frequency domain. The dynamic stiffness matrix is defined in terms of the shape functions matrix, $\bf{\Psi}(\dot{\imath}\omega)$, and the boundary conditions matrix, $\bf{G}(\dot{\imath}\omega)$, as follows 
\begin{eqnarray}
\label{eqn:EOM}
\mathbf{F} &=&\mathbf{K}(\dot{\imath}\omega)\mathbf{d} 
       =\mathbf{\Psi}(\dot{\imath}\omega)^{-1} \mathbf{G}(\dot{\imath}\omega)\mathbf{d},
\end{eqnarray}
where $\bf{d}$, $\bf{F}$, $\bf{\Psi}$, and $\bf{G}$ for a two-node spectral finite element are defined in Appendix A.

Although spatial discretization is not required for uniform, homogeneous structures, three spectral finite elements are used for the current analysis. This is done to allow the application of the desired excitation configuration described in Figure \ref{fig:beam}. Standard assembly procedures are applied to obtain the global dynamic stiffness matrix.

The beam considered here is a $48$-in. long rectangular beam, with a  $1\times1/8-in.^2$ cross sectional area. The beam is made of Aluminum, with $E= 69~GPa$, $G=26~GPa$ and $ \rho= 2700~kg/m^3$. Piezoelectric wafers are bonded to the beam's upper and lower surfaces, $18.5~in.$ from its left end. A total of $23$ receptance \frfs~are calculated. These are the driving-point \frf~{(calculated at the right edge of the piezoelectric actuator)} along with $22$ transfer \frfs~calculated over a span of $22$ in., with  $1$-in. increments. \frfs~are calculated over the frequency range of $0$ to $50~kHz$ with $0.25~Hz$ resolution. Several boundary conditions are investigated in this study. The analysis is first presented for the free-free case. Other boundary conditions are discussed in Section \ref{BCs}. 

\section{Data-driven Rational Approximation}
\label{sec:VF}
This section briefly outlines the  mathematical framework employed to construct rational approximations (state-space models) from a given set of \frf~measurements. To make the notation more clear and the discussion more concise, the methodology for only scalar measurements is presented, i.e., it is assumed that \frf~measurements come from a single-input single-output dynamical system. Theory and numerical implementations are already established for multi-input multi-output systems, and the reader is referred to \cite{drmac2015vector,gustavsen1999rational} for those details. 

 Let $\Hf(\dot{\imath} \omega) \in \mathbb{C}$  denote the 
 \frf~of the underlying dynamics evaluated at the frequency $\omega$. Then, the starting point is a set of \frf~samples at selected frequencies, denoted by  $\{\Hf( \dot{\imath} \omega_1),\Hf( \dot{\imath} \omega_2), \ldots,\Hf( \dot{\imath} \omega_N)\}$ where $N$ is a positive integer. In this paper where the theoretical framework is established, these \frfs~are obtained by sampling the analytical transfer functions derived from Eq. \eqref{eqn:EOM}.
More specifically, the components of the inverse of the dynamic-stiffness matrix, i.e., $\mathbf{K}^{-1}(\dot{\imath}\omega)$, corresponding to the out-of-plane and in-plane displacements are sampled over the frequency range of interest. In practice, these samples will be obtained via experimental measurements.

Given the samples, $\{\Hf( \dot{\imath} \omega_j)\}$ for $j=1,2,\ldots,N$, the goal is, now, to construct a degree-$r$ dynamical system described by its degree-$r$ rational function (\frf), $\Hr(s) = \ssCr(s I - \ssAr)^{-1}\ssBr$, that fits the data in an appropriate sense. The degree-$r$ means that 
$\ssAr \in \mathbb{R}^{r \times r}$, 
 $\ssBr \in \mathbb{R}^{r \times 1}$, and 
 $\ssCr \in \mathbb{R}^{1 \times r}$. 
 In this work, the least-squares (\ls) measure is used to judge the quality of the data-driven model: 
  \begin{equation} \label{lsprob}
       \mbox{Find~the~degree-$r$~rational~function~} \Hr(s) \mbox{~that~minimizes~}
 \sum_{j=1}^N \left| \Hf(\dot{\imath} \omega_j) - \Hr(\dot{\imath} \omega_j) \right|^2.
  \end{equation}

The  Vector Fitting (\vf) method of Gustavsen et al. \cite{gustavsen1999rational} is used (and adopted to our setting) in solving the \ls~problem \eqref{lsprob}. This approach is briefly explained in this section.

Let $n(s) = \alpha_0 + \alpha_1 s + \cdots+ 
\alpha_{r-1} s^{r-1}$ and $d(s) = \beta_0 + \beta_1 s + \cdots+ 
\beta_{r-1} s^{r-1}+ s^r$ denote, respectively, the numerator and denominator of 
the degree-$r$ rational function $\Hr(s)$, i.e.,  
$$\Hr(s) = \frac{n(s)}{d(s)} =
\frac{\alpha_0 + \alpha_1 s + \cdots+ 
\alpha_{r-1} s^{r-1}}{\beta_0 + \beta_1 s + \cdots+ 
\beta_{r-1} s^{r-1}+ s^r}.
$$  
The numerator $n(s)$ has degree-$r-1$ (as opposed to degree-$r$) since $\Hf(s)$ does not have a direct feed through term; thus, it is a strictly proper rational function. The formulation can be easily modified to allow a degree-$r$ numerator; but for the problems presented herein, $\Hf(s)$ is strictly proper. Now notice that the unknowns appear both in $n(s)$ and $d(s)$; thus the minimization problem \eqref{lsprob} is a \emph{nonlinear} \ls~problem and cannot be solved with direct methods in one step, as is the case for linear  \ls~problems. An iterative scheme is needed: the nonlinear \ls~problem  \eqref{lsprob} is solved iteratively by solving a sequence of linear \ls~problems. This is achieved by linearizing the error function $\Hf(s) - \Hr(s)$. Using ${\displaystyle \Hr(s) = \frac{n(s)}{d(s)}}$, the \ls~error in \eqref{lsprob} can be  rewritten as 

\begin{equation}
\sum_{j=1}^N \left| \Hf(\dot{\imath} \omega_j) - \Hr(\dot{\imath} \omega_j) \right|^2 = 
\sum_{j=1}^N \frac{\left|d(\dot{\imath} \omega_j) \Hf(\dot{\imath} \omega_j) - n( \dot{\imath} \omega_j)\right|^2}{\left| d(\dot{\imath} \omega_j)\right|^2},
\end{equation} 
which is still nonlinear in the variables. Starting with the $d^{(0)}(s) \equiv 1$, 
Sanathanan and Koerner  \cite{Sanathanan-Koerner-1963} proposed an iterative scheme where at the $k^{th}$ step
\begin{equation}\label{eq:relaxedNLS}
\mbox{~the~\ls~error}~\displaystyle \sum_{k=1}^N  \left| \frac{n^{(k+1)}(\xi_i) - d^{(k+1)}(\xi_i)H(\xi_i)}{d^{(k)}(\xi_i)} \right|^2~\mbox{is~minimized~by~solving~for~}
n^{(k+1)}(s)~\mbox{and}~d^{(k+1)}(s).
\end{equation}
Note the difference from the earlier problem. The relaxed \ls~problem in \eqref{eq:relaxedNLS} is linear in the variables $n^{(k+1)}(s)$ and $d^{(k+1)}(s)$, and  can be solved using linear \ls~solution techniques. Then, the process is repeated using the new denominator  $d^{(k+1)}(s)$ until convergence is achieved. This is the \textsf{SK} iteration to solve the original nonlinear \ls~problem \eqref{lsprob}.

The \vf~method is a further improvement on the \textsf{SK} iteration and uses a different parameterization for $\Hr(s)$; specifically, \vf~uses the barycentric-form for a rational function: At the $k^{th}$ step of the \vf~iteration, the approximation is parameterized as

\begin{equation}\label{eq:H_r(k)}
\Hr^{(k)}(s) =
\frac{\tilde{n}^{(k)}(s)}{\tilde{d}^{(k)}(s)} =
\frac{\sum_{j=1}^r \phi_j^{(k)}/(s-\lambda_j^{(k)})}{1 + \sum_{j=1}^r \psi_j^{(k)}/(s-\lambda_j^{(k)})},\;\;\mbox{~where~~}
\phi_j^{(k)},\psi_j^{(k)}, \lambda_j^{(k)} \in\mathbb{C}.
\end{equation}

The advantage of this formulation is that the \emph{poles} $\lambda_j^{(k)}$ are an arbitrary set of mutually distinct points. Note that $\lambda_j^{(k)}$ are \emph{not} the poles of $\Hr^{(k)}(s)$ unless $\psi_j^{(k)} = 0$ for $j=1,2,\ldots,r$. As in the \textsf{SK} iteration, for fixed poles $\lambda_j^{(k)}$, the \ls~problem for $\Hr^{(k)}(s)$ in \eqref{eq:H_r(k)} becomes linear in the remaining variables $\phi_j^{(k)}$ and $\psi_j^{(k)}$. Then, the \vf~algorithm continues the iteration by updating the poles $\lambda_j^{(k)}$ as the zeros of the numerator in Eq. \eqref{eq:H_r(k)}, i.e., as the zeroes of $ \tilde{d}^{(k)}(s) = 1 + \sum_{j=1}^r \psi_j^{(k)}/(s-\lambda_j^{(k)})$.  This iteration is run until convergence of the poles $\lambda_j^{(k)}$,  upon which the denominator  $\tilde{d}^{(k)}(s) \to 1$ and the final rational approximant is given by

\begin{equation}
\Hr(s) = \sum_{j=1}^r \frac{\phi_j}{s-\lambda_j} =
\ssCr(s I - \ssAr)^{-1}\ssBr.
\end{equation}
For further details, we refer the reader to \cite{gustavsen1999rational,drmac2015quadrature}.
\vf~has successfully been used in many applications and various modifications have been proposed. The pole relocation step has been improved \cite{gustavsen2006improving}, the MIMO case has been efficiently parallelized \cite{Chinea-G.Talocia-2011}, it has been combined with quadrature-based sampling \cite{Drmac-Gugercin-Beattie:VF-2014-SISC}, and it has been analyzed in detail in terms of numerical robustness \cite{drmac2015vector}. The reader is referred to \cite{berljafa2017rkfit} where a rational Krylov toolbox has been developed to solve similar nonlinear \ls~problems. The Loewner framework, introduced by Mayo and Antoulas \cite{mayo2007fsg}, is another commonly used approach to construct rational approximants from \frf~measurements. In this case, the approximant is sought to be an (approximant) interpolant as opposed to a \ls~fit. The recently developed Adaptive Anderson-Antoulas algorithm \cite{nakatsukasa2018aaa} is a hybrid approach where a rational interpolant is constructed to interpolate a subset of the data and to minimize \ls~error in the rest. As pointed out earlier, in this work, the focus is on the \ls~framework where the regular \vf~framework is used to solve the data-driven modeling problem.

\section{Data-driven Model based Dispersion Curves for Flexural Waves}
\label{Method}
The proposed approach for estimating dispersion curves utilizes \frfs~to generate a single-input multi-output (SIMO), data-driven, state-space model, $\Hfit$, of the structure under consideration. This model is then used to simulate the transient dynamic response of the structure to a given excitation. Simulated responses are then analyzed in the frequency domain to reconstruct dispersion curves. The analysis is first presented for out-of-plane (flexural) receptance \frfs, corresponding to the first anti-symmetric wave mode. The analysis is repeated for the first symmetric wave mode, using the in-plane (longitudinal) \frfs, in Section \ref{in-plane}. 
The process of estimating dispersion curves from \frfs~can be split into two main stages: (i) the development of the data-driven model using \vf~and (ii) transient response simulation and analysis. The two stages are discussed in detail in the following subsections. 
\subsection{Data-driven Modeling using \vf}
\label{Subsec:VF}
In this section, a SIMO, data-driven, state-space model, $\Hfit$, is developed based on the simulated \frfs~for the beam over the frequency range of $0.01$ to $50$ $kHz$.
\frfs~are sampled such that the frequency resolution is $0.25 Hz$ over the entire frequency range of $50kHz$. Such fine frequency resolution is needed to better capture lower-order resonant frequencies,
resulting in approximately 200,000 samples. As discussed in detail below, an accurate $\Hfit$ for the full frequency spectrum requires an approximation order of $r>200$. Since we are measuring $23$ \frfs, this means that every \vf~iteration requires solving a linear least-squares problem with a dense coefficient matrix having more than $2.5\times 10^6$ rows and $4800$ columns. Since 
the coefficient matrix changes at every step, this is not a trivial numerical task and needs careful numerical implementation such as those in \cite{Chinea-G.Talocia-2011,drmac2015vector}.

However, the main issue in trying to fit the \frfs~directly on the full frequency spectrum is the sensitivity to the initial pole selection. As discussed in Section \ref{sec:VF}, the \vf~algorithm is initialized by a set of poles. While for modest $r$ values and for a smaller number of measurements, the \vf~algorithm does a good job of correcting the poles during the iteration, for the scale of the problems and for the complex dynamics considered here, a not-well-informed initial pole selection leads to inaccurate rational approximants together with potentially highly ill-conditioned least-squares problems to solve; in other words, for the setting considered in this paper, the quality of the fitted model is more dependent on the choice of starting poles; see  \cite{gustavsen1999rational,gustavsen2006improving,berljafa2017rkfit} for related work. Moreover, when experimentally measured noisy \frfs~are used, these issues will be further magnified. Therefore, in order to develop a framework that is much better suited for experimentally collected data, a more comprehensive initialization procedure is developed for \vf.  For this purpose, the full frequency range is divided into multiple smaller bands. Since high-quality rational approximations in smaller frequency bands can be obtained via smaller approximation orders, for each frequency band, the initial, but now much smaller number of  poles for \vf, are selected by taking into account the quantity and the location of resonant frequencies. {Additionally, due to the free-free boundary conditions, rigid-body modes appear in the \frfs~at $0~Hz$. While the \vf~ algorithm can handle fitting these modes, numerical instabilities arising from the $0~Hz$ pole renders simulation results inaccurate. To avoid such numerical instabilities, rigid-body modes are eliminated in the current analysis. }

 Once these smaller frequency bands are fitted, the resulting poles are put together to initialize the \vf~algorithm for the entire frequency range. With this much improved initial pole selection, \vf~converges quickly for the whole frequency range and yields an accurate rational approximation, as illustrated in the numerical examples below. The following steps further discuss the details of this process.

\begin{steps}[start=1]
\item \textbf{\vf~of \frf~Partitioned into Multiple Bands}
\end{steps}
 Since for dispersive media, such as beams and plates, the frequency spacing between resonant frequencies is not uniform, random assignment of initial poles would not be practical. {For instance, in 1-dimensional dispersive structures, the  frequency spacing between consecutive natural frequencies increases with mode number. As a result, modal density is higher at lower frequency ranges, i.e., there are more resonant peaks in the $0-1~kHz$ range than, for example, the $2-3~kHz$ range.}  This becomes even more important when dealing with noisy experimental data. As a result, the \vf~algorithm is initialized using poles that are obtained by fitting \frfs~over smaller frequency bands.   

{When fitting out-of-plane \frfs, the full frequency range is divided into smaller bands in order to achieve high-quality rational approximations. Seven frequency bands, with comparable modal density, were selected for this purpose. Details of the number of resonant peaks and the corresponding number of poles used to fit \frf~in each band is tabulated in Table \ref{tab:tab_section}.} Each resonant peak of the \frf~is represented by a pair of complex-conjugate poles, where the $n^{th}$ pair of poles are of the form $p^{2n-1} = \alpha +i\beta$, $ p^{2n} = \alpha-i\beta$. Therefore, the number of poles of the fitted model are at least twice the number of resonant frequencies in the frequency band of interest. While the imaginary part ($\beta$) of these poles indicates the oscillatory behavior, the real part ($\alpha$) represents the damping characteristic of the dynamical system. Resonant peak frequencies are used as the initial values for $\beta$, whereas the real part is initialized as $\alpha$ $=$ $\beta/100$, following the recommendation of \cite{gustavsen1999rational}. The effect of initial pole placement and model order on the performance of this approach will be further investigated in future studies. Table \ref{tab:tab_section} summarizes fitting results for each frequency band. 
The quality of the fit is determined using a relative $L_2 $ error defined  as 
\begin{eqnarray}
\relerr &=& 
\frac{1}{N}\sqrt{
{ \frac{{\displaystyle\sum_{i=1}^N}\norm{\Hf(\dot{\imath} \omega_i) - \Hr(\dot{\imath} \omega_i)
}_F^2}{{\displaystyle\sum_{i=1}^N} \norm{\Hf(\dot{\imath} \omega_i)}_F^2} }}.
\end{eqnarray}
where $\| \cdot \|_F$ denotes the Frobenious norm. Table \ref{tab:tab_section} demonstrates that the constructed models fit the data to a high relative accuracy.

\begin{figure}[ht!]
\centering
\includegraphics{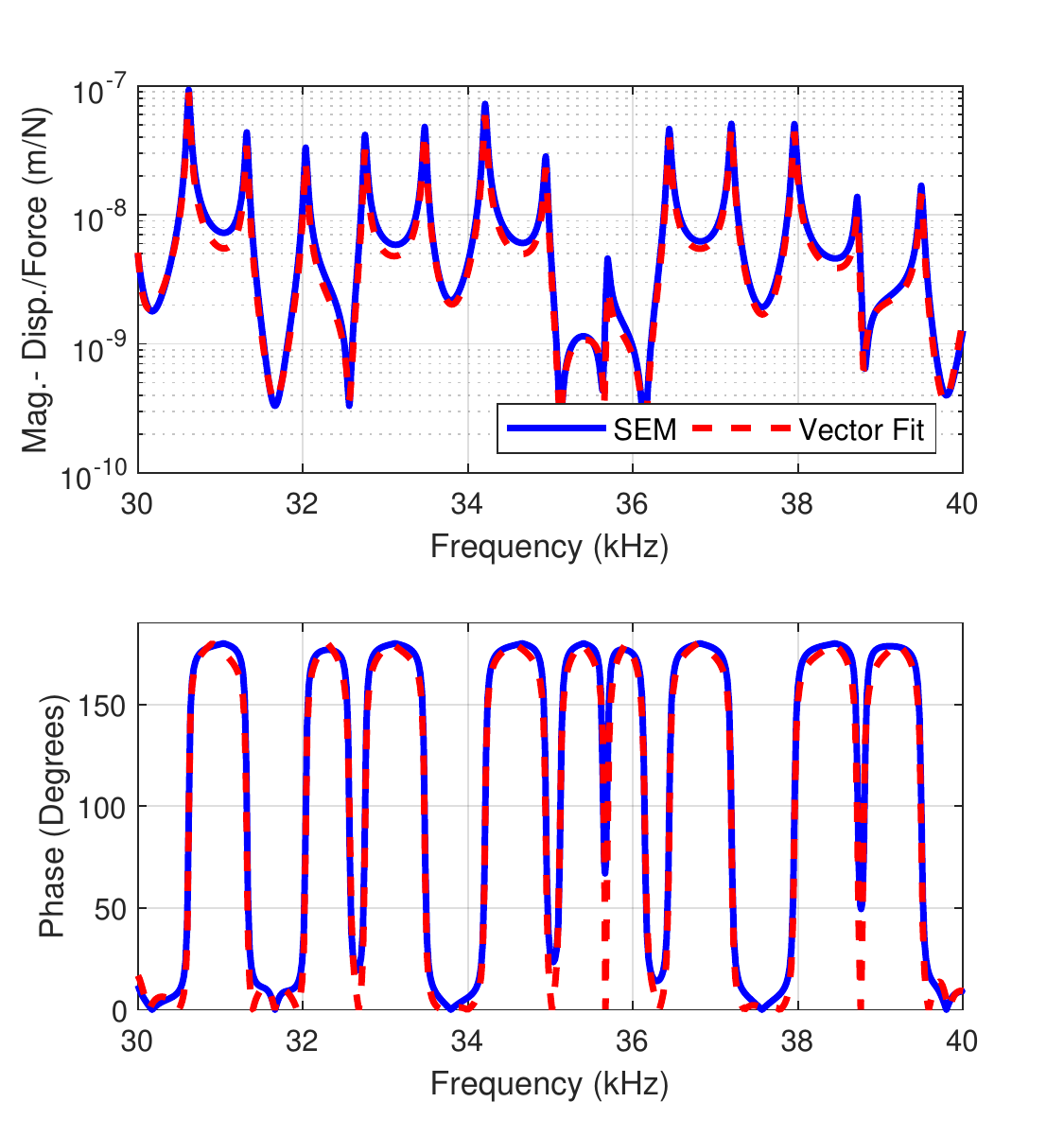}
\caption{Out-of-plane receptance \frfs~of the beam obtained by the data-driven state space model compared to the SEM predictions over the frequency range of $30~kHz$ to $40~kHz$. \frfs~are calculated $1~in.$ away from the excitation point.}
\label{fig:FRF_fit}
\end{figure}

While the minimum number of poles required to fit each frequency band is twice the number of resonant peaks in that band, adding extra poles to account for near by resonant peaks, that have strong influence on the \frfs~in that band, was found to be necessary. For example,  at least $28$ complex poles were needed to fit the \frf~within the frequency band of $0$ - $1~kHz$, which only has $13$ resonant peaks. A closer examination of the fitted poles shows that the two additional poles correspond to a resonant peak at $1038~Hz$, which is very close to the current frequency band of interest. Table \ref{tab:tab_section} shows the number of resonant peaks in each frequency band along with the number of poles needed to fit \frfs~over each band.
\begin{table}[h]
\centering
\caption{Number of resonant peaks and initial poles in partitioned frequency bands}
\label{tab:tab_section}
\begin{tabular}{ @{}c|c|cc |cc@{} } 
 \toprule
   Band & Frequency range &  Resonant peaks & No. of poles &$\relerr$ \\ [0.5ex]
  \hline \rule{0pt}{2.2ex}
 1 & $ 0.01$ - $1kHz$  & 13  & 28 & $1.02\times10^{-5} $ \\ 
   2 & $ 1$ - $5kHz$  & 18  & 40& $3.78\times10^{-6} $  \\ 
   3 & $5$ - $10kHz$  & 13  & 32& $2.96\times10^{-6} $  \\ 
   4 & $10$ - $20kHz$  & 21  & 44& $1.48\times10^{-6} $  \\ 
   5 & $20$ - $30kHz$  & 16  & 38& $1.45\times10^{-6} $  \\ 
   6 & $30$ - $40kHz$  & 13  & 30& $1.44\times10^{-6} $  \\ 
   7 & $40$ - $50kHz$  & 12  & 26& $1.42\times10^{-6} $  \\
   \bottomrule
   \end{tabular}
\end{table}
\begin{steps}[start=2]
\item \textbf{\vf~of the Entire \frf}
\end{steps}
Once the poles of the state-space models for each frequency band are determined, they are combined together to initialize \vf~over the entire frequency range. While combining these poles, it is important to avoid duplicating them. The complete \frf~with 106 resonant peaks is fitted with 212 poles, exactly the minimum number of poles required to fit all resonant peaks over this frequency range. The result is a data-driven, state-space model of the form:
\begin{eqnarray}
\Hfit &=& \ssCr(s\bm{I}-\ssAr )^{-1}\ssBr,  \label{eq2}
\end{eqnarray}
where the state matrices are $\ssAr \in  {\rm I\!R}^{212 \times 212}$, $\ssBr \in  {\rm I\!R}^{212 \times 1}$, and $\ssCr \in  {\rm I\!R}^{23 \times 212}$.  Figure \ref{fig:FRF_fit} depicts the magnitude and phase of the fitted \frf~($\Hr(\dot{\imath} \omega)$) as compared to the out-of-plane component of the  \frf~matrix ($\bf{K}^{-1}(\dot{\imath} \omega)$),  derived from Eq. \eqref{eqn:EOM} using the SEM, 
for the frequency band of $30~kHz$ to $40~kHz$. Resonant peaks and anti-resonant valleys are accurately captured by the fitted model. It should be noted that an inverse weighting scheme is adopted with the least square fit during the \vf~process. Thus, the \vf~algorithm places more importance on fitting the anti resonances than the resonant peaks. This is particularly important as the intended application of the data-driven model is to simulate the transient response of the system over the entire frequency range.
The relative $L_2$ error, $\relerr$, for individual frequency bands is provided in the Table \ref{tab:tab_section} and the error for the complete frequency range is $\relerr ~=~ 2.13\times10^{-7}$.

\subsection{Transient Response Analysis and Dispersion Curves Estimation}
Once the data-driven, state-space model
$\Hfit$ in Eq. \eqref{eq2}
is developed, the transient response of the structure under test can be numerically approximated by 
exciting $\Hfit$ with tone-burst input signals. Dispersion curves are then estimated by analyzing the waveforms propagating along the structure. Details of the this process are further presented through the following steps: 
\begin{steps}[start=1]
\item \textbf{Transient Response Simulations}
\end{steps}
The transient response of the structure under test is simulated using the developed data-driven model. An amplitude-modulated, sine wave, tone burst is selected as the excitation waveform. This waveform is selected as it is found to minimize the separation distance between the measurement location and structure's boundaries required for obtaining a reflection-free response \cite{albakri2016novel}. In this study, the number of cycles is selected to  be $2$ when the strongly dispersive flexural waves are being considered and $1$ for the weakly dispersive longitudinal waves.

\begin{figure}[!ht]
    \begin{minipage}[h]{0.5\linewidth}\centering
  \includegraphics{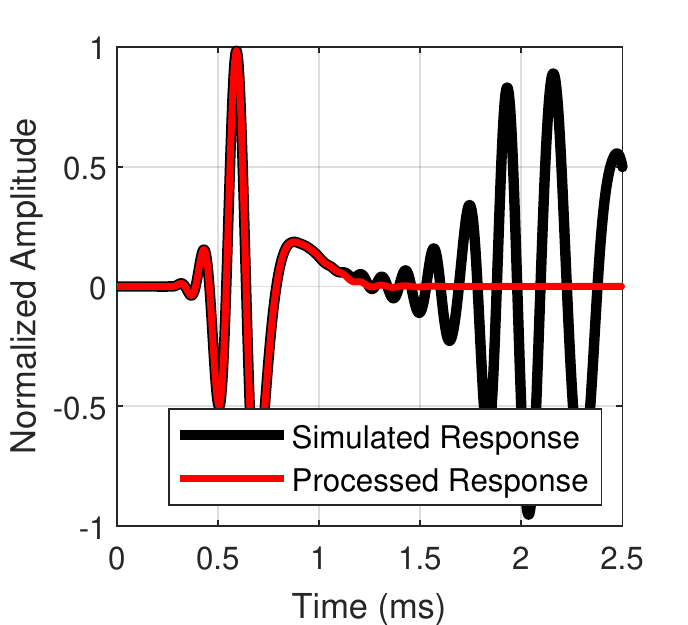}\\
	(a)
  \end{minipage}
  \hfill
  \begin{minipage}[h]{0.5\linewidth}\centering
  \includegraphics{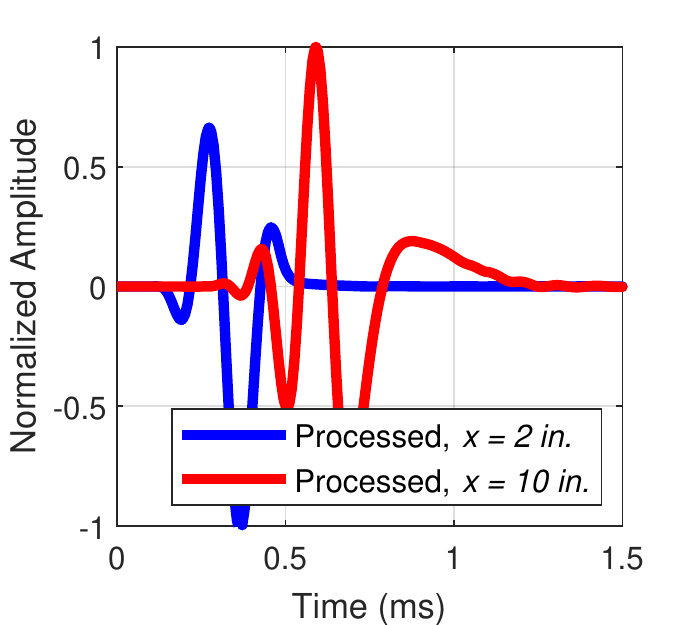}\\
	(b)
  \end{minipage}
  \caption{Flexural waveforms showing (a) simulated and processed responses $10~in.$ away from the excitation point and (b) processed responses $2~in.$ and $10~in.$ away from the excitation point. Responses are simulated with a $2$-cycle sine wave tone burst excitation signal, with $5~kHz$ central frequency.}
	\label{fig:Out_response}
\end{figure}

The data-driven model is then used to calculate the response of the structure at all 22 points on the beam where \frfs~are originally obtained.  Figure \ref{fig:Out_response}.a shows the simulated response $10~in.$ away from the excitation point as a $2$-cycle, sine wave, tone burst of a central frequency of $5~kHz$ is applied. As noted in the figure, due to the finite dimensions of the beam, reflections from boundaries is present in the simulated response. Structure's boundary conditions, free-free in this case, determine how the incident wave is reflected. This information is carried by the \frfs~that have been used to obtain the data-driven model. Boundary condition effects on estimating dispersion curves are further discussed in Section \ref{BCs}.

\begin{steps}[start=2]
\item \textbf{Incident Waveform Extraction and Processing}
\end{steps}
Once the transient response is simulated at a given location, the incident waveform is extracted. This is performed by determining the first dominant peak in the simulated response, defining a time-window containing the incident wave, and then applying a strongly decaying exponential functions outside that window. The size of this window is determined based on the number of cycles in the excitation signal, excitation frequency, and the location where the response is simulated. Figure \ref{fig:Out_response}.a shows the extracted incident wave, labeled \textit{Processed Response} as compared to the original simulated response, $10~in.$ away from the excitation point. Figure \ref{fig:Out_response}.b shows the processed responses $2~in.$ and $10~in.$ away from the excitation point. The dispersive nature of the flexural waves at this frequency range is evident in the figure where waveform distortion and spread can be noticed as the wave propagates along the beam.

\newpage
\begin{steps}[start=3]
\item \textbf{Dispersion Curves Estimation}
\end{steps}

Given the \textit{Processed Response} at locations $i$ and $i+1$, the wavenumber, $k$, corresponding to the wave mode under consideration can be calculated as follows
\begin{eqnarray}
\label{eqn:Propagation}
\mathbf{U_{i+1}}\left ( \omega \right)=\mathbf{U_i}\left ( \omega \right ) e^{-i  \mathbf{k} \left ( x_{i+1}-x_i \right)},
\end{eqnarray}
where $\mathbf{U_i}$  is the vector of Fourier coefficients of the signal measured at location $x_i$. At the frequency range of interest, the only wave mode contributing to the out-of-plane receptance \frfs~is the first anti-symmetric mode. Thus, the wavenumber vector, $\mathbf{k}$, only includes that wave mode. Equation \eqref{eqn:Propagation} is solved for $\mathbf{k}$ and the group velocity is then calculated as $V_G=\partial \omega / \partial k$. The process is then repeated where the central frequency of the excitation wave is varied to sweep the frequency range of interest. {Equation \eqref{eqn:Propagation} requires the response at two distinct points to be known. With the response being simulated at 23 points in this study, 253 different combinations can be obtained, and hence, 253 estimates of the dispersion curves can be calculated. This allows for statistical analysis to be conducted and confidence bands to be defined for the estimated dispersion curve. Such analysis is important when dealing with experimental data where noise and measurement errors can be present. Simulated responses, on the other hand, do not suffer from such sources of error, thus statistical analysis has not been included in this study.}

\begin{figure}[ht!]
\centering
\includegraphics{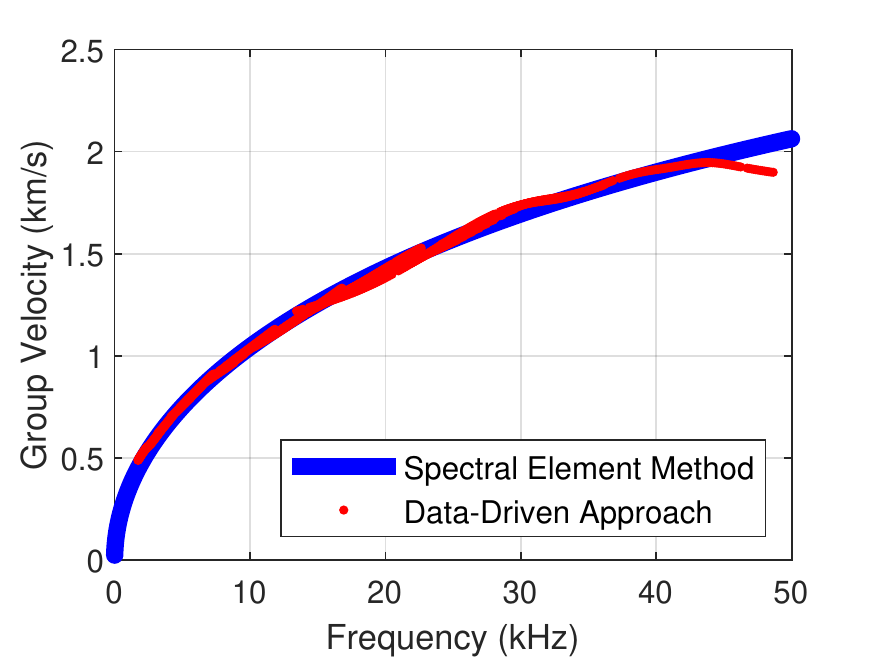}
\caption{Comparison of group velocity curves estimated using the proposed data-driven approach (red) and that predicted by the SEM (blue) for the first anti-symmetric (flexural) wave mode.}
\label{fig:Dispersion_FF}
\end{figure}

Figure \ref{fig:Dispersion_FF} depicts the group velocity of the flexural (the first anti-symmetric) wave mode calculated using the proposed data-driven modeling approach as compared to that predicted by the SEM. As depicted in the figure, the results of the data-driven modeling approach accord very well with the SEM predictions. It should be noted that at high frequency, greater than $45~kHz $, the data-driven model fails to predict the dispersion curve. This can be ascribed to the fact that the \frfs~that have been used to obtain the data-driven model over the frequency range of $0-50~kHz$. Since a $2-$cycle sinusoidal waveform is relatively broad-band, high-frequency signals of this form have frequency content that extends well beyond the $50~kHz$ limit of the model. {Increasing the number of cycles in the excitation tone burst reduces the bandwidth of the excitation signal. This allows dispersion curves to be correctly estimated at frequencies closer to the measurement ceiling. However, the model remains limited by the frequency range over which \frfs~have been obtained.}

\FloatBarrier

Close examination of Figure \ref{fig:Dispersion_FF} also reveals that the very low-frequency part of the dispersion curve is missing. This is due to the large wavelength of flexural waves at such low frequencies compared to the length of the beam under test. This, in turn, hinders the separation of incident and reflected  waveforms, a prerequisite for Eq.\eqref{eqn:Propagation} to be applicable. While this is an inescapable limitation for conventional wave-propagation-based experimental techniques, the use of the proposed data-driven approach allows for a number of solutions to be applied. {One such solution can be to introduce artificial damping to the data-driven model in order to attenuate reflected waves. This will be the focus of future studies.}

The results presented in this section highlight the capabilities of the proposed approach where a model derived from the steady-state dynamic response of the structure is employed to accurately capture its transient response and estimate its dispersion curves. The following section extends this discussion to the longitudinal wave mode.

\section{Data-driven Modeling based Dispersion Curves for Longitudinal Waves}
\label{in-plane}
In this section, the applicability of the proposed approach to calculate the dispersion curves of the first symmetric (longitudinal) wave mode is investigated. The process presented in this section closely follows the one discussed in Section \ref{Method}, the main difference is that in-plane receptance \frfs~of the beam are utilized to obtain the data-driven model.

\begin{figure}[ht!]
\centering
\includegraphics{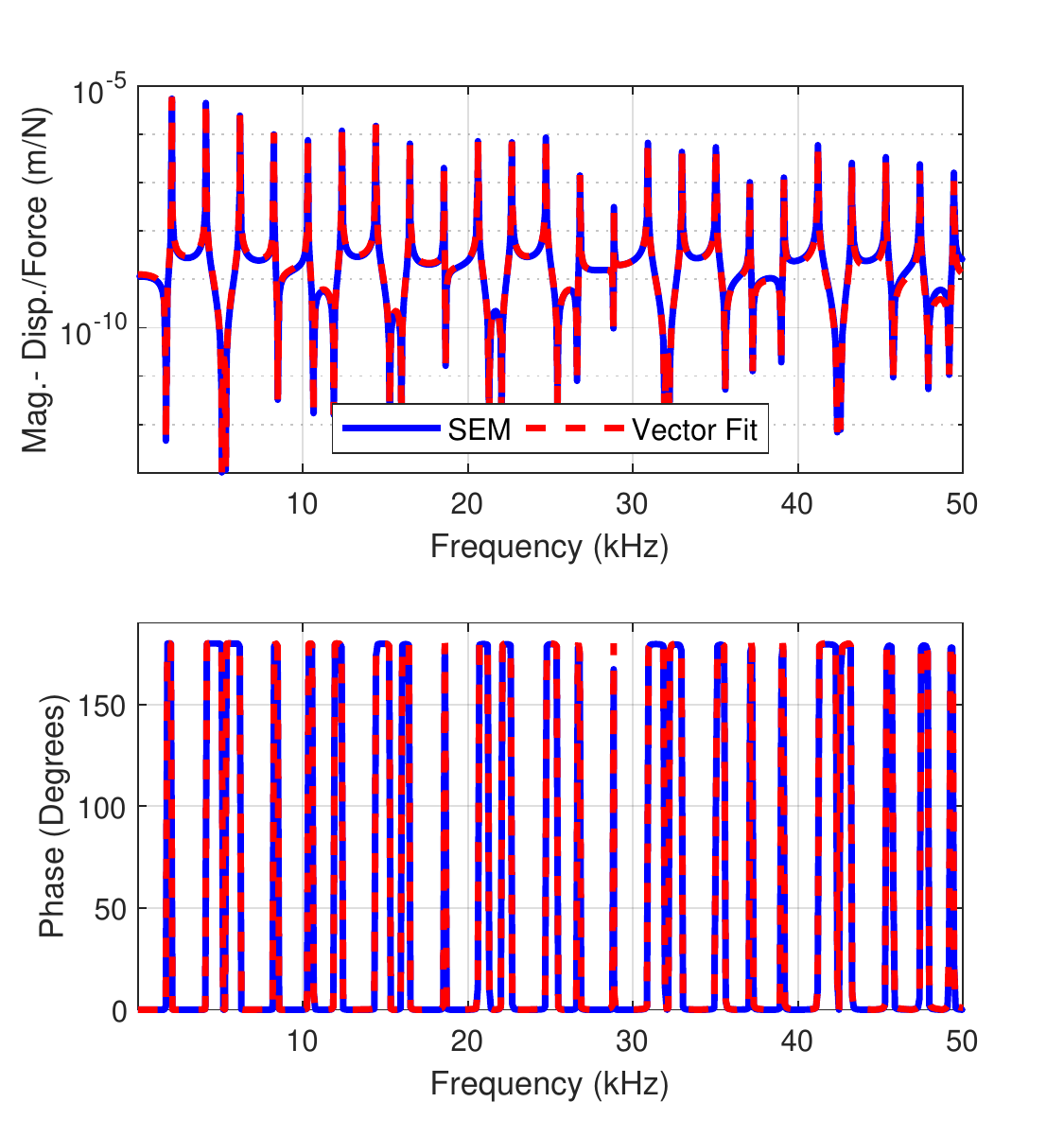}
  \caption{In-plane receptance \frf~of the beam obtained by the data-driven state space model compared to the SEM predictions. \frfs~are calculated $6~in.$ away from the excitation point over the frequency range of $0-50~kHz$.}
  \label{fig:In_plane_FRF}
\end{figure}

In-plane \frfs~are obtained by exciting the upper and lower faces of the beam in-phase, as shown in Figure \ref{fig:beam}. Since the beam under test is perfectly symmetric with respect to its neutral axis, in-plane and out-of-plane deformation are completely uncoupled. Thus, the aforementioned excitation configuration results in pure longitudinal deformations without any contribution from flexural deformations. Figure \ref{fig:In_plane_FRF} shows the transfer, longitudinal, receptance \frf~, $6-in$ away from the excitation location, as obtained by the SEM. As is the case for the out-of-plane deformations, the analysis is limited to the frequency range of $0-50~kHz$.

\begin{table}[h]
\centering
\caption{Fitting results and errors for the various data-driven models used in this study.}
\label{tab:error_BCs}
\def\arraystretch{1.5}
\begin{tabular}{ @{}c|cc|cc@{} } 
 \toprule
   Boundary Conditions  && State Matrices &$\relerr$ \\ [0.5ex]
  \hline
 Free - Free    && $\ssAr \in  {\rm I\!R}^{212 \times 212}$, $\ssBr \in  {\rm I\!R}^{212 \times 1}$, $\ssCr \in  {\rm I\!R}^{23 \times 212}$ & $2.13\times10^{-7} $ \\
 Clamped - Free  && $\ssAr \in  {\rm I\!R}^{214 \times 214}$, $\ssBr \in  {\rm I\!R}^{214 \times 1}$, $\ssCr \in  {\rm I\!R}^{16 \times 214}$ & $2.75\times10^{-7} $ \\ 
 Pinned -Pinned   && $\ssAr \in  {\rm I\!R}^{212 \times 212}$, $\ssBr \in  {\rm I\!R}^{212 \times 1}$, $\ssCr \in  {\rm I\!R}^{16 \times 212}$ & $3.09\times10^{-7} $ \\
  Free - Free (inplane)   && $\ssAr \in  {\rm I\!R}^{48 \times 48}$, $\ssBr \in  {\rm I\!R}^{48 \times 1}$, $\ssCr \in  {\rm I\!R}^{23 \times 48}$ & $3.12\times10^{-7} $ \\

   \bottomrule
   \end{tabular}
\end{table}
Following the procedure outlined in Section \ref{Subsec:VF}, \vf~is utilized to fit the simulated longitudinal \frfs~of the beam. In the frequency range of interest, the $24$ resonant peaks are fitted with $48$ complex-conjugate poles. The nondispersive nature of the longitudinal waves is evident from the harmonic nature of the resonant peaks. Given the relatively small number of resonant peaks in this frequency range, \frf~partitioning was unnecessary in this case. Table \ref{tab:error_BCs} presents the order of the fitted state-space matrices and the corresponding $\relerr$ error between the fitted \frfs~and the simulated ones. Additionally, the quality of the fit is illustrated by comparing the fitted \frf~to the SEM prediction as shown in Figure \ref{fig:In_plane_FRF}.

\begin{figure}[ht!]
  \begin{minipage}[h]{0.5\textwidth}\centering
  \includegraphics	{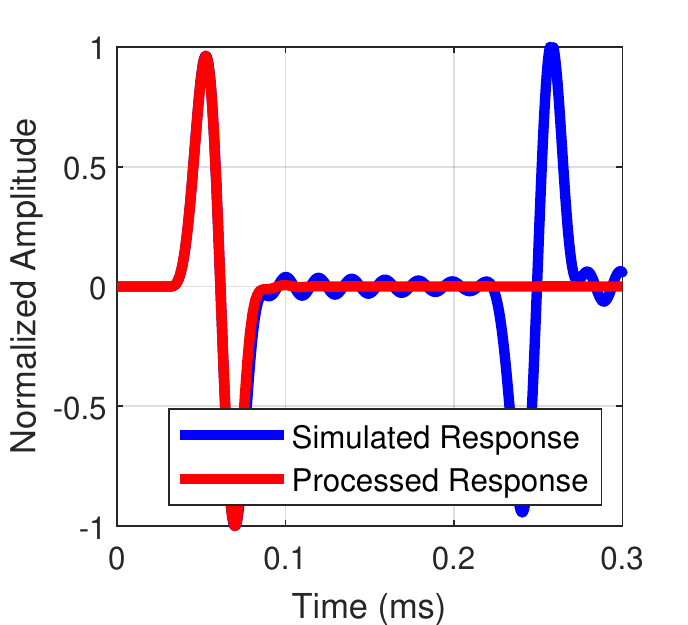}\\
	(a)
  \end{minipage}
  \hfill
  \begin{minipage}[h]{0.5\textwidth}\centering
  \includegraphics	{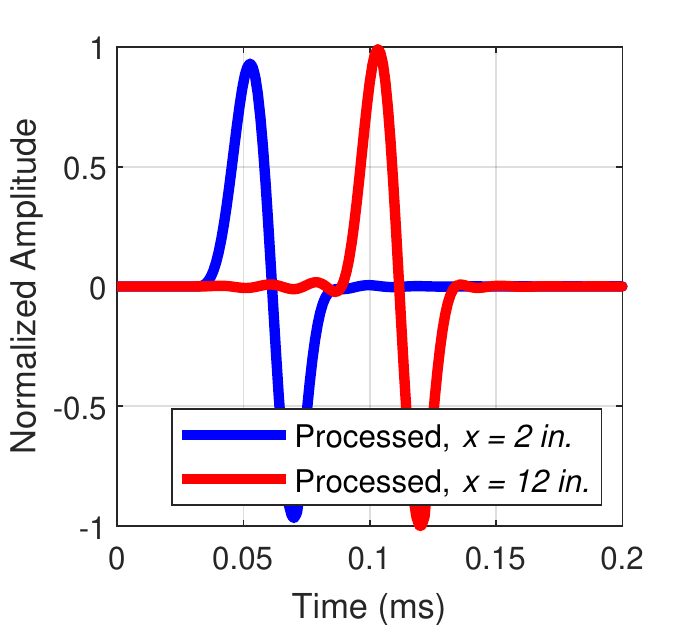}\\
	(b)
  \end{minipage}
  \caption{ Longitudinal waveforms showing (a) simulated and processed responses $2~in.$ away from the excitation point, and (b) processed responses $2~in.$ and $12~in.$ away from the excitation point. Responses are simulated with a $1$-cycle sine wave tone burst excitation signal, with $20~kHz$ central frequency.}
  \label{fig:Inpalne_response}
\end{figure}

The data-driven state-space model is then used to simulate the transient response of the system. As the longitudinal wave mode is nondispersive at this frequency range, a $1$-cycle sine-wave tone burst is used as the excitation signal. Incident waveforms are then extracted from the simulated response. Figure \ref{fig:Inpalne_response}.a shows the extracted incident wave, labeled \textit{Processed Response} as compared to the original simulated response $2~in.$ away from the excitation point. Figure \ref{fig:Inpalne_response}.b shows the processed responses $2~in.$ and $12~in.$ away from the excitation point. The nondispersive nature of this wave mode is seen in the figure where waveforms propagate along the beam without noticeable distortion.

\begin{figure}[ht!]
\centering
\includegraphics{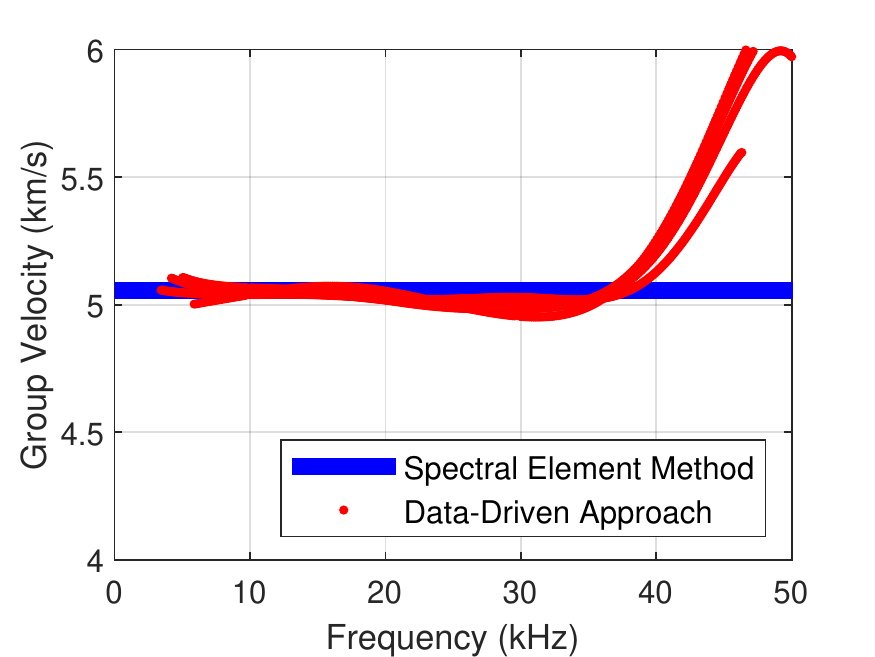}
\caption{Comparison of group velocity curves estimated using the proposed data-driven approach (red) and that predicted by the SEM (blue) for the first symmetric (longitudinal) wave mode.}
\label{fig:Dispersion_inplane}
\end{figure}

While simple time-of-flight calculations can be used to estimate wave speed of nondispersive wave modes, the frequency-domain analysis presented in the previous section is followed. With the frequency-domain analysis, a given waveform can be used to estimate wave speed over a relatively wide frequency band, as opposed to a single frequency estimate with the time-of-flight calculations. Thus, statistical techniques can be implemented to define confidence intervals for the estimated dispersion curves, which is important when dealing with experimental measurements. Following Eq. \eqref{eqn:Propagation}, the processed longitudinal waveforms at locations $i$ and $i+1$ are analyzed and the wavenumber, $k$, corresponding to the first symmetric wave mode is obtained for each frequency in the excitation signal. Group velocity is then calculated and the results are shown in Figure \ref{fig:Dispersion_inplane}. As depicted in the figure, the group velocity calculated using the data-driven approach accord very well with the SEM predictions. textcolor{red}{At high frequency,$ >40~kHz$, the proposed approach fails to predict the dispersion curve. This is due to the broadband nature of the $1-$ cycle tone bursts and the limited frequency range over which \frfs~are obtained.} The very low-frequency part of the dispersion curve is also missing since the large wavelength of longitudinal waves at such low frequencies hinders the separation of incident and reflected  waveforms.\\

\section{On the Effects of Boundary Conditions}
\label{BCs}
\begin{figure}[ht!]
\centering
\includegraphics {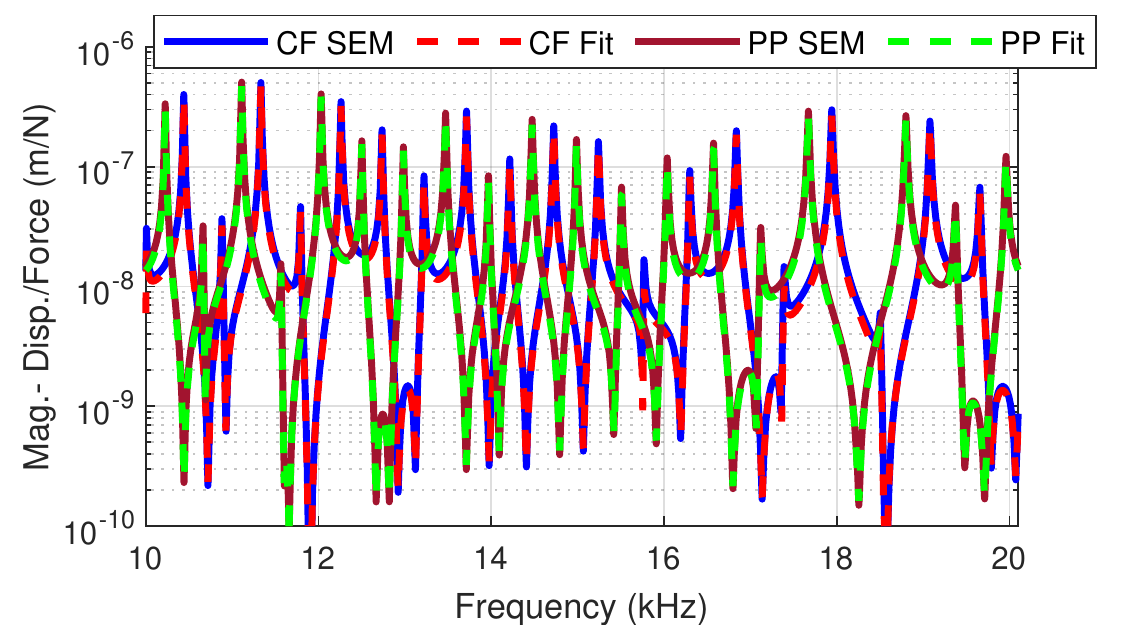}
\caption{Out-of-plane receptance \frfs~of the beam with clamped-free and pinned-pinned boundaries over the frequency range of $10~kHz$ to $20~kHz$. Simulated and fitted \frfs~are compared. \frfs~are calculated $1~in.$ away from the excitation point.}
\label{fig:Fit_BCs}
\end{figure}

In this section, the effects of boundary conditions on the performance of the proposed approach are investigated. While end conditions do not affect wave propagation characteristics along the waveguide, they determine how waves are reflected at the boundaries, which has a profound impact on measured \frfs. The goal of this section is to demonstrate that such changes in \frfs~will not affect dispersion curves estimates. For this purpose, two additional combinations of  boundary conditions are investigated, these are clamped-free and pinned-pinned conditions. out-of-plane receptance \frfs~for these boundary conditions are obtained using the SEM. The \vf~algorithm is then used to create a data-driven, SIMO model for each set of boundary conditions. Table \ref{tab:error_BCs} summarizes fitting results for these boundary conditions, along with the free-free conditions discussed in the previous sections. The dimensions of the state matrices and the relative fit error for each case are given in the table. Figure \ref{fig:Fit_BCs} depicts out-of-plane receptance \frfs~of the beam with clamped-free and pinned-pinned boundary conditions, where simulated and fitted \frfs~are compared. The impact of boundary conditions on \frfs~is evident. 

\begin{figure}[ht!]
  \begin{minipage}[h]{0.5\textwidth}\centering
\includegraphics[]{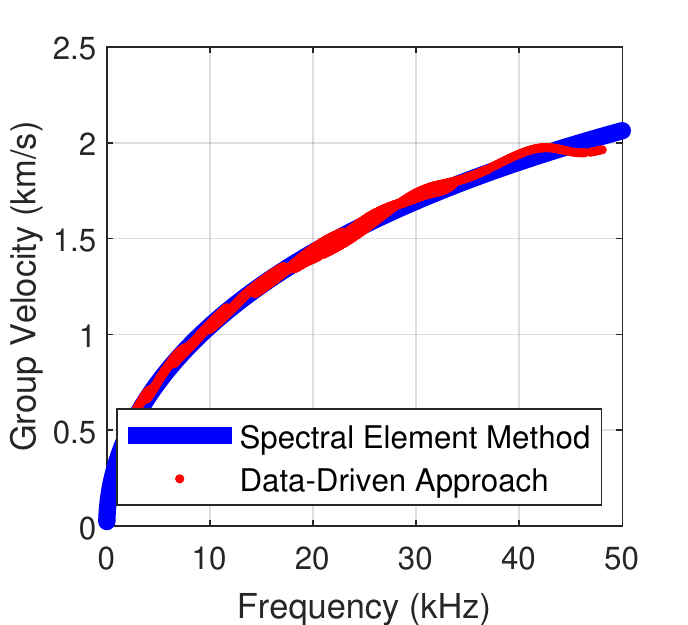}\\
	(a)
  \end{minipage}
\begin{minipage}[h]{0.5\textwidth}\centering
 \includegraphics[]{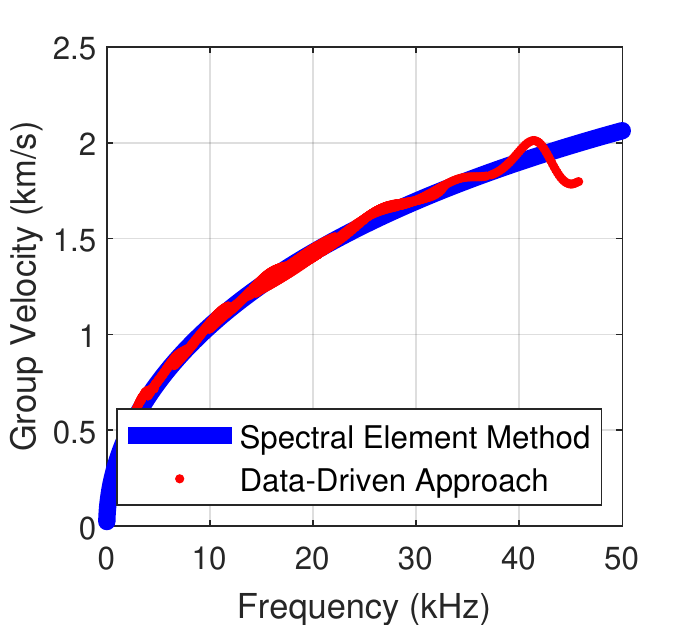}\\
	(b)
  \end{minipage}
  \caption{Group velocity curves for the first anti-symmetric wave mode estimated using (a) fixed-free, and (b) pinned-pinned data-driven models.}
	\label{fig:Dispersion_BCs}
\end{figure}

Following the procedure outlined in Section \ref{Method}, dispersion curves are estimated using the clamped-free, and the pinned-pinned, data-driven models. The results are summarized in Figure \ref{fig:Dispersion_BCs}. Although the \frfs~used to obtain the data-driven models are affected by changes in boundary conditions, such effects are not reflected on the estimated dispersion curves. Both clamped-free and pinned-pinned models are capable of accurately estimating dispersion curves, as suggested by the figure. It should be noted that the previously discussed limitations at the upper- and lower-ends of the frequency range are applicable to the current cases. While the analysis presented in this section is limited to flexural deformations, the conclusions are equally valid for longitudinal deformations. 
{Observations based on the numerical experiments presented in this section highlight the capability of the approach in dealing with various boundary conditions. In future studies, the robustness of the proposed approach to uncertainties and noise in the experimental data of structures under real-boundary conditions will be investigated. }

\section{Conclusions}
In this paper, a new data-driven modeling approach for estimating dispersion curves is presented. The Vector Fitting method is adopted to develop a single-input-multi-output, state-space, dynamical model of the beam under test based on receptance \frfs. \frfs~corresponding to longitudinal and lateral deformations of the beam are simulated over the frequency range of $0-50~ kHz$. The data-driven  model is able to accurately capture the dynamic behaviour of the system within the frequency range of interest with an error of \textbf{$\relerr = 2.13 \times 10^{-7}$} for lateral deformations, and  \textbf{$\relerr = 3.12 \times 10^{-7}$}  for longitudinal deformations.

The data driven model is then used to study wave propagation along the beam and obtain its dispersion curves. Group velocity curves corresponding to symmetric and anti-symmetric wave modes are estimated with this approach over the frequency range of $2-40~kHz$. The results are found to be in very good agreement with the numerical predictions of the SEM. Dispersion curves at frequencies lower than $2~kHz$ cannot be estimated directly using this approach. This is due to the large wavelength at such low frequencies compared to the length of the beam under test, which hinders the separation of incident and reflected waveforms. While this is a limitation of all current experimental practices, the proposed approach allows for potential solutions to be implemented, such as the introduction of artificial damping to attenuate reflected waves, which will be addressed in future studies. 

The effects of boundary conditions on the performance of the developed approach have also been studied in this work. Although boundary conditions have a profound impact on the \frfs~that are used for the data-driven models, estimated group velocity curves are found to be unaffected by such conditions. This allows for this approach to be used regardless of boundary conditions of the structure under test, which is crucial for many practical applications.

The work presented herein demonstrates the feasibility of estimating dispersion curves based on steady-state \frfs~using this new data-driven modeling approach. The experimental implementation of this technique will investigated in future studies, where the issues of model order selection, parameter identification, and the introduction of artificial damping will be investigated. {Future studies will also address the scenarios where different wave modes are coupled or simultaneously excited where the impact of coupling at the \frf-level on predicted dispersion curves will be investigated.}

\section*{Acknowledgment}
Tarazaga would like to acknowledge the support provided by the John R. Jones III Faculty Fellowship. The work of Gugercin was supported in parts by NSF through Grant DMS-1522616 and DMS-1720257. The work of Albakri was supported in parts by the Federal Railroad Administration.

\section*{Appendix A: Spectral Element Matrices}
For a two-node spectral finite element, vectors $\mathbf{d}$ and $\mathbf{F}$ in Eq. \ref{eqn:EOM} as defined as follows
\begin{eqnarray}
\mathbf{d}=\left \{ \begin{array} {cccccc} U_0 (x_1) & W_0 (x_1) & \Phi (x_1) & U_0 (x_2) & W_0 (x_2) & \Phi (x_2)\\ \end{array} \right \}^T \nonumber, \\
\mathbf{F}=\left \{ \begin{array} {cccccc} -\bar{F}_x(x_1) & -\bar{F}_z (x_1) & -\bar{M} (x_1) & \bar{F}_x(x_2) & \bar{F}_z (x_2) & \bar{M} (x_2)\\ \end{array} \right \}^T \nonumber,
\end{eqnarray}
where $U_0$, $W_0$, $\Phi$, $\bar{F}_x$, $\bar{F}_z$, and $\bar{M}$ are, respectively, the longitudinal displacement, the lateral displacement of the beam's neutral axis, the angle of rotation of the neutral axis normal vector, the externally applied axial force, lateral force, and bending moments in the frequency domain. $x_1$ and $x_2$ are the coordinates of the left and right nodes of the spectral finite element.\\

The shape functions matrix, $\mathbf{\Psi}(\dot{\imath} \omega)$, in Eq. \ref{eqn:EOM}, is defined for a two-node spectral finite element as follows
\begin{eqnarray*}
\mathbf{\Psi}(\dot{\imath} \omega)=\left [ \begin{array} {cccccc} 
\zeta_{11} & \zeta_{21} & 0 & 0 & 0 & 0\\
0 & 0 & \zeta_{31} & \zeta_{41} & \zeta_{51} & \zeta_{61}\\ 
0 & 0 & r_{33}\zeta_{31} & r_{34}\zeta_{41} & r_{35}\zeta_{51} & r_{36}\zeta_{61}\\
\zeta_{12} & \zeta_{22} & 0 & 0 & 0 & 0\\
0 & 0 & \zeta_{32} & \zeta_{42} & \zeta_{52} & \zeta_{62}\\ 
0 & 0 & r_{33}\zeta_{32} & r_{34}\zeta_{42} & r_{35}\zeta_{52} & r_{36}\zeta_{62}\\
\end{array} \right ],
\end{eqnarray*}
where $\zeta_{mj}=e^{-\dot{\imath}k_m\left (\dot{\imath} \omega \right) x_j}$ with $m=1,2,...,6$, and $j=1,2$.\\

The non-zero elements of the boundary conditions matrix, $G \left(\dot{\imath} \omega \right)$ are \begin{eqnarray*}
G_{1s}&=&\dot{\imath}k_{sn}EA e^{-\dot{\imath}k_{sn}x_1}\\
G_{2t}&=&GA\bar K \left(-\dot{\imath}k_{tn} +r_{3t}\right)e^{-\dot{\imath}k_{tn}x_1}\\
G_{3t}&=&-\dot{\imath}EIk_{tn}e^{-\dot{\imath}k_{tn} x_1}\\
G_{4s}&=&\dot{\imath}k_{{tn}sn}EA e^{-\dot{\imath}k_{sn}x_2}\\
G_{5t}&=&GA\bar K \left(-\dot{\imath}k_{tn} +r_{3t}\right)e^{-\dot{\imath}k_{tn} x_2}\\
G_{6t}&=&-\dot{\imath}EIk_{tn} e^{-\dot{\imath}k_{tn}x_2}
\end{eqnarray*}
where $s=1,2$ and $t=3,4,5,6$.

\bibliographystyle{elsarticle-num.bst}
\bibliography{Data_driven.bib}

\end{document}